\documentclass[10pt]{revtex4}
\usepackage{hyperref}
\usepackage{amsmath}
\usepackage[psamsfonts]{amssymb }
\usepackage{mathrsfs}

\begin{document}

\title{\Large Hamiltonian and Lagrangian BRST quantization in Riemann Manifold}
\author{Vipul Kumar Pandey\footnote {e-mail address: vipulvaranasi@gmail.com}}

\affiliation { Department of Physics and Astrophysics, 
University of Delhi, New Delhi, 110007, INDIA. }

\begin{abstract}
The BRST quantization of particle motion on the hypersurface $V_{(N-1)}$ embedded in Euclidean space $R_N$ is carried out both in Hamiltonian and Lagrangian formalism. Using Batalin-Fradkin-Fradkina-Tyutin (BFFT) formalism, the second class constrained obtained using Hamiltonian analysis are converted into first class constraints. Then using BFV analysis the BRST symmetry is constructed. We have given a simple example of these kind of system. In the end we have discussed Batalin-Vilkovisky formalism in the context of this (BFFT modified) system. 
\end{abstract}
\maketitle
\section{Introduction}
The quantum mechanical analysis of the system in curved space has been examined about the ordering problem for a long time. Primarily, two approaches have been used, canonical quantization and path-integral method \cite{LLSKOT,RS,TKM,TKRS,TORS1,TORS2,TKTORS,MOHS,FSTK,KFAK,TK,HKTK,JGANMO,MO,BS}. Also, the problem of the quantization of a dynamical system constrained to a curved manifold embedded in the higher-dimensional Euclidean space has been extensively investigated as one of the quantum theories on a curved space \cite{BDMOHS,VASFMR,NBD,LPDT,IBMGSL1,IBMGSL2,IGSLAS}. Here we are taking a non-relativistic particle constrained to a curved surface embedded in the higher dimensional Euclidean space \cite{OFK,OFCK}. These type of systems and their various properties such as quantization in different approaches and their comparisons has been studied by many authors \cite{NOG1,MIYN,STTT,CFFM,NOG2,FOUC,NCKFAK,ASA,DLLHQL,NOG3,NOG4,NOMN,DMOFGK,AFHGPK,AVG,KFNO,MNNOHM1,MNNOHM2,MN,PM1,PM2,NCAR1,NCAR2,NCAR3,SOCSRT}. Here for the first time we have explicitly constructed the BFFT Abelianization and BRST symmetry for the system of a non-relativistic particle constrained to a curved surface embedded in the higher dimensional Euclidean space in both Hamiltonian and Lagrangian formalism \cite{VKP2}. The results derived are of highly significant because of nonlinear nature of the constraints of the system. This is the first time BFFT Abelianization and BRST quantization of a nonlinear constrained system has been constructed explicitly. 

BRST quantization \cite{BRS1,BRS2,BRS3,IVT} is an important and powerful technique to deal with a field theoretic model with gauge symmetry. It has also been found as symmetry of constrained systems \cite{PAMD1,JAPB,PAMD2,MH,KS,MHCT}. It enlarges the phase space of a gauge theory and restores the symmetry of the gauge fixed action in the extended phase space without changing the physical contents of the theory. BRST symmetry plays a very important role in renormalizing spontaneously broken theories, like standard model and hence it is extremely important to investigate it for different systems. To the best of our knowledge BRST formulation for a non-relativistic particle constrained to a curved surface embedded in the higher dimensional Euclidean space which is nonlinear constrained system which is a toy model for a wide class of physical systems has not been developed yet. This motivates us in the study of BRST symmetry for this system. We study non-relativistic particle constrained to a curved surface embedded in the higher dimensional Euclidean space following the technique of Dirac constraints analysis. The system is shown to contain second-class constraints which are nonlinear in nature. We will apply Batalin-Fradkin-Fradkina-Tyutin (BFFT) method to convert these second class constraints to first class constraints \cite{IBEF1,IBEF2,IBEFTF,IBIT,EERM,RBJBN,CBAPMT,IBVKAR,ILBAAR,AAR,IBAR,VPRT}. We will further develop the BFV (Batalin-Fradkin-Vilkovisky) formulation of this BFFT extended theory using the constraints of the theory \cite{EFGV,IABGV,IBEF} by the Faddeev-Senjanovic technique \cite{LDFAD,PSENJ}. The nilpotent BRST charge is constructed in the operator form using the mode expansion of the fields \cite{MKKOSH,SH}. The result has been verified using simple example of particle on torus \cite{STH,VPBM,DXQLXZ,RK}. At the end we will construct BRST transformation of the system using Batalin-Vilkovisky (BV) quantization \cite{IABGAV,IBGV,JGJPSS,RART}.  
                         This is the first part of the two part manuscript. In the second part we will discuss BRST quantization of embedding $V_L$ in Euclidean space $R_N$ where $1\leq L < N$ \cite{VKP}. This manuscript has been arranged in the following way. In the first section, we have reviewed motion on hypersurface (H.S.) and its equivalence with motion in curved space and also calculated all the possible constraints of the theory using Dirac's constraints analysis. In the second section, we have reviewed BFFT formalism. In the third section we have constructed first class constraints and Hamiltonian. In the next section we have constructed BRST symmetry for the system based on BFV Formalism. We have also constructed BRST operator. In the fifth section we have given a simple mechanical example of this kind of system. In the sixth section, we have discussed BV quantization of this system based on BFFT formalism. In the last section concluding remark has been made.
 
\section{Motion on a hypersurface: a review}

Consider an N dimensional Euclidean space ${R_N}$, which is specified by a set of Cartesian coordinates
                         
                         $\{x_1,x_2,...,x_a,...,x_n\}$. Further
consider in $R_N$ the $(N-1)$ dimensional H.S., $V_{(N-1)}$ subject to the equation
      $f(x_1,x_2,...,x_N) = 0$. Let us consider the motion of particle on this H.S. with potential $V(x)$ \cite{OFK,OFCK}.    
The Lagrangian for this system can be written as,
\begin{eqnarray}
L_0 = \frac{1}{2}\cdot{\dot x}^a{\dot x}_a - V(x) + \lambda f(x)
\label{Lagz}
\end{eqnarray}
here, $\delta_{ab}$ is the metric, $\lambda$ is a variable, which is independent of $x_a (a;1,2,...,N)$ and the dot denotes the time derivative.
The canonical momentum conjugate to $x^a$ and $\lambda$ can be written as
\begin{eqnarray}
P_a &=& \frac{\partial L}{\partial {\dot x}^a} = {\dot x}^a \nonumber\\
P_\lambda & = & \frac{\partial L}{\partial {\dot \lambda}} \approx 0
\label{conmom}
\end{eqnarray}
Hamiltonian corresponding to Lagrangian in eqn(\ref{Lagz}) can be written as,
\begin{eqnarray}
H _0 = \frac {1}{2}\cdot P_a P^a + V(x) - \lambda f(x)
\label{zoham}
\end{eqnarray}

\subsection{Equivalence with Particle in curved space }
Let us consider a general coordinate transformation of $N$ dimensional Euclidean space \cite{OFK}.
\begin{eqnarray}
x^a &\longrightarrow & q^\mu\nonumber\\
a:&1\sim N& \quad \mu : 0\sim (N-1)
\label{xaq1N}
\end{eqnarray}
In this coordinate frame $q^i$'s are the coordinates on H.S. and $q^0$ is the coordinate normal to it. In this coordinate frame, constraints will take form 
\begin{eqnarray}
f(x) &\leftrightarrow & q^0 = \mathrm{constant}\nonumber\\
{\dot x}^a\partial_a f(x) &\leftrightarrow &{\dot x}^a\partial_a q^0 = {\dot q}^a = 0
\end{eqnarray}
Here we take the constant $q^0$ equal to zero.
So, the constraints to the particle motion on H.S. are,
\begin{eqnarray}
q^0 = 0, \quad {\dot q}^a = 0
\end{eqnarray}
Here $dq^i$ is a tangential vector, and $dq^0$ is the normal vector. The metric for $dq^\mu$ is generally given by,
\begin{eqnarray}
ds^2 = \sum_{a} dx^a dx^a = dx^a dx_a = g_{\mu\nu}dq^{\mu}dq^{\nu}
\end{eqnarray}
where metric $g_{\mu\nu}$ is defined as 
\begin{equation}
[g_{\mu\nu}] = 
\left[
\begin{array}{cc}
g_{00} & 0 \\
0 & g_{ij}\\
\end{array}
\right]
\label{gmunu}
\end{equation}
Here $g_{i0} = g_{0i} = 0$ which implies that $dq^0$ is normal to the Surface.
The inverse of the metric $g_{\mu\nu}$ is defined as
\begin{equation}
[g^{\mu\nu}] = {[g_{\mu\nu}]}^{-1} =
\left[
\begin{array}{cc}
g^{00} & 0 \\
0 & g^{ij}\\
\end{array}
\right]
\label{invgmn}
\end{equation}
Here, 
\begin{eqnarray}
g^{00} = \frac{1}{g_{00}}, and \quad g^{ij}.g_{jk} = {\delta^i}_k
\end{eqnarray}
Unit normal in Cartesian coordinate is defined \cite{OFK}
\begin{eqnarray}
n^a = n_a = \frac{\frac{\partial f(x)}{\partial x^a}}{[\partial_b f(x).\partial^b f(x)]^\frac{1}{2}}
\end{eqnarray}
Under a general coordinate transformation, $n^a$ is transformed into $n^\mu$ as, 
\begin{eqnarray}
n^\mu &=& \frac{\partial q^\mu}{\partial x^a}.n^a,\nonumber\\
n_\mu &=& g_{\mu\nu}.n^\nu = \frac{\partial x^a}{\partial q^{\mu}}.n_a
\end{eqnarray} 
In terms of metric, unit normal can be defined as
\begin{eqnarray}
n^\mu &=& {\delta^\mu}_0(g^{00})^{\frac{1}{2}},\nonumber\\
n_\mu &=& {\delta^0}_\mu(g_{00})^{\frac{1}{2}}
\end{eqnarray}
Using the transformation in general coordinate system discussed above, we will find the equations in modified form as:
Equation of motion:
\begin{eqnarray}
{\ddot q}^i + \Gamma^i_{jk}{\dot q}^j {\dot q}^k + g^{ik}.\frac{v(q)}{\partial q^k} = 0\nonumber\\
{\ddot q}^0 = 0
\end{eqnarray}
constraints:
\begin{eqnarray}
q^0 = {\dot q}^0 = 0
\end{eqnarray}
Here potential is defined as
\begin{eqnarray}
V(x) = V(x(q)) = V(q).
\end{eqnarray}
The equations derived above implies that there is no motion along the normal and the equation of motion is quite the same as Euler-Lagrange equation obtained from Lagrangian $L= \frac{1}{2}\cdot g_{ij}{\dot q}^i {\dot q}^j - V(q)$ which implies that classically the equation of motion in a curved space is similar to that of on HS. In the general coordinate system, the form of Hamiltonian derived from Lagrangian is written as:
\begin{eqnarray}
H = \frac{1}{2}.g^{ij}(q) p_i p_j + V(q)
\end{eqnarray}

\subsection{Hamiltonian Analysis} 
The primary constraint for this system is
\begin{equation}
P_{\lambda} \approx 0
\end{equation} 
After inclusion of primary constraint our new Hamiltonian has the form
\begin{equation}
H_T = \frac {1}{2}\cdot P_a P^a + V(x) - \lambda f(x) + u P_\lambda
\end{equation}
where $u$ is the Lagrange multiplier.
Now, using the Dirac's technique of constraint analysis \cite{PAMD1,JAPB,PAMD2,KS,MHCT,MH}, we will calculate all the constraints of the theory \cite{OFK}.
\begin{eqnarray}
&&{\dot P}_\lambda =  \{ P_\lambda, H_T \}_P = f(x) \approx 0 \nonumber\\
&&{\ddot  {P}}_\lambda  =  \{ f(x), H_T \}_P = P^a \cdot \frac{\partial f(x)}{\partial x^a} \approx 0 \nonumber\\
&&{P_\lambda}^{(3)} = \{ P^a\cdot \frac{\partial f(x)}{\partial x^a}, H_T \}_P = P^a P^b{\partial_a}{\partial_b}f(x) \nonumber\\ &&- (\partial_b V(x) - \lambda\partial_b f(x))\cdot \partial^b f(x)\approx 0 
\end{eqnarray}
${P_\lambda}^{(4)}$ will vanish and the value of $u$ will be determined from it. All the constraints can be written as,
\begin{eqnarray}
\Phi_1 &=& P_\lambda \approx 0 \nonumber\\
\Phi_2 &=& f(x) \approx 0 \nonumber\\
\Phi_3 &=& Df(x) \approx 0 \nonumber\\
\Phi_4 &=& D^2 f(x) - {\partial_b}(V - \lambda f(x))\cdot \partial^b f(x) = D^2 f(x) \nonumber\\ &&- {\partial_b}\Phi\cdot \partial^b f(x) \approx 0 
\label{cnoth}
\end{eqnarray}
where $D = P^a \partial_a$ and $\Phi = (V - \lambda f(x))$.
Now, the Poisson brackets between the constraints have following values,
\begin{eqnarray}
\{\Phi_1, \Phi_4 \}_P &=& - \partial_a f(x)\cdot \partial^a f(x) = - \alpha \nonumber\\
\{\Phi_2, \Phi_3 \}_P &=& \partial_a f(x)\cdot \partial^a f(x) = \alpha \nonumber\\
\{\Phi_2, \Phi_4 \}_P &=& P^a\partial_a (\partial_b f(x)\cdot \partial^b f(x)) = P^a\partial_a \alpha = -\beta \nonumber\\
\{\Phi_3, \Phi_4 \}_P &=& 2\partial_a (D f(x))\cdot \partial^a (D f(x)) - \partial_a f(x)\nonumber\\&&\cdot \partial^a (D^2 f(x) - {\partial_b}\Phi\cdot \partial^b f(x)) = - \gamma 
\end{eqnarray}
Thus the matrix $\Delta_{ab}$ between the constraints has the form
\begin{equation}
\Delta_{ab} = \{\Phi_a, \Phi_b\}_P =
\left[
\begin{array}{cccc}
0 & 0 & 0 & - \alpha \\
0 & 0 & \alpha & -\beta \\
0 & -\alpha & 0 & -\gamma \\
\alpha & \beta & \gamma & 0
\end{array}
\right]
\label{Delab}
\end{equation}

\section{BFFT Analysis: A short Review}
In this section we will review BFFT technique \cite{IBEF1,IBEF2,IBEFTF,IBIT,EERM,RBJBN,CBAPMT,IBVKAR,ILBAAR,AAR,IBAR}, which is used to construct a first class constraint system from a second-class constraint system. 
                              We know from the Dirac-Bergmann constraint
analysis that second-class constraint of a constrained system satisfy an open algebra. Let us take a system described by a Hamiltonian $H_0$ in a 2N dimensional phase space. Let us denote the second-class constraints of the system as $T_a$ with $a = 1,2,...,M < 2N$. These constraints satisfy following algebra
\begin{eqnarray}
\{ T_a, T_b\} = \Delta_{ab},
\label{secal}
\end{eqnarray}
where $\det (\Delta_{ab}) \neq 0$.
To achieve this goal, we will extend the Hilbert space of the theory by introducing auxiliary fields $\eta^a$, one for each second class constraint. This is done to keep the physical degrees of freedom in the extended theory same as in the original theory. These fields satisfy the symplectic algebra,
\begin{eqnarray}
\{ \eta_a, \eta_b \} = \omega^{ab}
\label{pofex}
\end{eqnarray}
where $\omega^{ab}$ is a constant quantity and $\det(\omega^{ab}) \neq 0$. The constraints are now defined in terms of auxiliary field $\eta_a$ as 
\begin{eqnarray}
{\tilde T}_a = {\tilde T}_a (q, p; \eta ) = \sum_{n = 0}^{\infty} {{\tilde T}_a}^{n} =  \sum_{n = 0}^{\infty}X_{(ab_{(n)})\gamma_{(n)}}
\eta^{(b_{(n)})\gamma_{(n)}}
\label{invcst}
\end{eqnarray}
The modified constraints satisfy first class constraints algebra. So the Poisson bracket between the constraints are defined as
\begin{eqnarray}
\{{\tilde T}_a, {\tilde T}_b\} = 0
\label{mfcal}
\end{eqnarray}
This modified constraint satisfies the boundary condition
\begin{eqnarray}
{\tilde T}_a(q, p; 0)  =  T_a (q, p) = {{\tilde T}_a}^{(0)},
\label{bdccn}
\end{eqnarray}
Replacement of eqn(\ref{invcst}) into eqn(\ref{mfcal}) gives recurrence relations, one for each coefficient of $\eta^n$. 

Using this iterative technique we can calculate the $n^{th}$ order correction term ${\tilde T}^{(n)}$.
The expression for ${\tilde T}^{(1)}$ is written as

\begin{eqnarray}
{\tilde T}_a^{(1)} = X_{ab}(q,p)\eta^b
\label{exton}
\end{eqnarray}
Putting this expression in (\ref{invcst}) and using the boundary condition (\ref{bdccn}) as well as (\ref{secal}) and (\ref{pofex}), we get
\begin{eqnarray}
\Delta_{ab} + X_{ac}\omega^{cd}X_{bd} = 0
\label{exton}
\end{eqnarray}
We notice that this equation does not give $X_{ab}$ univocally, because it also contains the still unknown $\omega_{ab}$. We choose $\omega_{ab}$ in such a way that the new variables are unconstrained. The knowledge of $X_{ab}$ allows us to obtain ${{\tilde T}_a}^{(1)}$. If $ T_a + {{\tilde T}_a}^{(1)}$ is strongly involutive then series ends here or we will continue the same process to calculate the higher order terms till we don't get strongly involutive constraints.

Another point in the Hamiltonian formalism is that any dynamic function $A(q, p)$ (for instance, the Hamiltonian) has also to be properly modified in order to be strongly involutive with the first-class constraints ${\tilde T}_a$. Denoting the modified quantity by $A(q,p;\eta)$, we then have \cite{BEF1,IBEF2,IBEFTF,IBIT,RBJBN}

\begin{eqnarray}
\{{\tilde T}_a, {\tilde A}\} = 0
\label{mHam}
\end{eqnarray}
In addition, ${\tilde A}$ has also to satisfy the boundary condition,
\begin{eqnarray}
{\tilde A}(q, p; 0)  =  A (q, p) = {\tilde A}^{(0)} 
\label{bdcHm}
\end{eqnarray}
To obtain $\tilde A$ an expansion analogous to (\ref{invcst}) is considered,
\begin{eqnarray}
{\tilde A} = \sum_{n = 0}^{\infty} A^{n}
\label{mdcHm}
\end{eqnarray}

$A^{(1)}$ can be written as
\begin{eqnarray}
A^{(1)} = - \eta^a\omega_{ab}X^{bc}(q,p)\{ T^c, A \},
\label{forHm}
\end{eqnarray}
where $\omega_{ab}$ and $X^{ab}$ are the inverses of $\omega^{ab}$ and $X_{ab}$.
It was earlier seen that $T^a + {T^a}^{(1)}$ was strongly involutive if the coefficients $X_{ab}$ do not depend on $(q, p)$. However, the same argument does not necessarily apply in this case. Usually we have to calculate other corrections to obtain the final $\tilde A$. 

The general expression reads as,
\begin{eqnarray}
A^{(n+1)} = - \frac{1}{n+1}\eta^a\omega_{ab}X^{bc}(q,p)G_c^{(n)},
\label{nocHm}
\end{eqnarray}
where
\begin{eqnarray}
G_a^{(n)} = \sum_{m = 0}^{n}\{T_a^{(n-m)}, A^{(m)} \}_{(q,p)} + \sum_{n = 0}^{n-2}\{T_a^{(n-m)}, A^{(m+2)}\}_{(\eta)} + \{T_a^{(n+1)}, A^{(1)}\}_{(\eta)}
\end{eqnarray}
Similarly the involutive form of other variables can be obtained using the method described above. Let the initial fields be $q$ and $p$. Then their involutive form $\tilde q$ and $\tilde p$ will follow these relations.
\begin{eqnarray}
\{{\tilde T}, {\tilde q}\} = \{{\tilde T}, {\tilde p}\} = 0
\label{minfl}
\end{eqnarray}
Now any function of $\tilde q$ and $\tilde p$ will also be strongly involutive, since
\begin{eqnarray}
\{{\tilde T}, {\tilde F}({\tilde q}, {\tilde p})\} = \{{\tilde T}, {\tilde q}\}\frac{\partial {\tilde F}}{\partial {\tilde q}} + \{{\tilde T}, {\tilde p}\}\frac{\partial {\tilde F}}{\partial {\tilde p}} = 0
\label{minfl}
\end{eqnarray}
Thus if we take any dynamical variable in the original phase space, its involutive form can be obtained by the replacement
\begin{eqnarray}
F (q,p) \rightarrow F({\tilde q}, {\tilde p}) = \tilde F ({\tilde q}, {\tilde p})
\end{eqnarray}
It is obvious that the initial boundary condition in the BFFT process, namely, the reduction of the involutive function to the original function when the new fields are set to zero, remains preserved.

\section{Construction of first class constraint Theory}
As all the constraints of the theory (eqn(\ref{cnoth})) are second class, we will introduce four possible fields $\vartheta^1, \vartheta^2, \vartheta^3, \vartheta^4$ corresponding to each constraint. Relation between these fields will give us possible solution of the eqn(\ref{exton}). Relation between these fields will provide us possible value of $\omega^{ab}$. Here we will discuss the solutions (based on author's recent article on Abelianization of prototypical nonlinear second class system \cite{VPRT}) and will construct the involutive Hamiltonian. 
Our choice of Poisson Bracket between the fields $\vartheta^1, \vartheta^2, \vartheta^3, \vartheta^4$ are
\begin{eqnarray}
\{ \vartheta^1, \vartheta^3 \} &=& 1, \quad \{ \vartheta^2, \vartheta^4 \} = 1
\label{btfl1}
\end{eqnarray}

From the above relation, matrix $\omega^{ab}$ between the fields can be written as,
\begin{equation}
\omega^{ab} = 
\left[
\begin{array}{cccc}
0 & 0 & 1 & 0 \\
0 & 0 & 0 & 1 \\
-1 & 0 & 0 & 0 \\
0 & -1 & 0 & 0
\end{array}
\right]
\label{omgab1}
\end{equation}
Putting the matrix $\omega^{ab}$ and $\Delta_{ab}$ in the eqn(\ref{exton}), and solving it, one can find many possible values of matrix $X_{ab}$. Now using the eqn(\ref{invcst}) we can calculate the modified constraints as
\begin{eqnarray}
\tilde\Phi_1 &=& P_\lambda - \vartheta^{(3)}\nonumber\\
\tilde\Phi_2 &=& f(x) + \vartheta^{(2)} \nonumber\\
\tilde\Phi_3 &=& (\bar P^a - {\partial^a {\bar f(x)}}\vartheta^{(4)})\bar{\partial_a f(x)}  \nonumber\\
\tilde\Phi_4 &=& (\bar P^a - \partial^a \bar f(x)\vartheta^{(4)})(\bar P^b - \partial^b \bar f(x)\vartheta^{(4)})\partial_a \partial_b \bar f(x) - \partial_a \bar V\cdot \partial^a \bar f(x) \nonumber\\ &&+ \lambda \partial_a \bar f(x)\partial^a \bar f(x) + \partial_a \bar f(x)\partial^a \bar f(x)\vartheta^1
\label{modcn1}
\end{eqnarray}
where all barred quantities are function of $x^a$ and $\vartheta^2$ and will change to original unbarred quantities in the limit $\vartheta^2 \rightarrow 0$. Here, any field $\bar A(x^a, \vartheta^{(2)})$ will be written as \cite{VPRT}
\begin{eqnarray}
{\tilde A (x^a, \vartheta^{(2)})} = \sum_{n = 0}^{\infty} \frac{A^{(n)}}{n!}\vartheta^{(2)}_{n}
\label{baxth}
\end{eqnarray}
also, its partial differentiation wrt. any field $x^a$ can be written as \cite{VPRT}
\begin{eqnarray}
{\tilde A_{,i}} = f(x)_i\{\bar A, \vartheta^{(4)}\}
\label{baxth}
\end{eqnarray}
The Poisson bracket between these modified constraints are
\begin{eqnarray}
\{{\tilde\Phi}_i, {\tilde\Phi}_j\} = 0
\label{evcnt1}
\end{eqnarray}
where $i, j = 1,2,3,4$.
which shows that modified constraints are involutive. Hence we have converted the second class constraints of the theory into first class.
Now, we will construct first class Hamiltonian for this system. 
Corrections in Hamiltonian due to different fields $\vartheta$ can be calculated as follows.
First we will calculate inverse of the matrices $\omega^{ab}$ and $X_{ab}$. The matrix $\omega_{ab}$ is written as
\begin{equation}
\omega_{ab} =
\left[
\begin{array}{cccc}
0 & 0 & -1 & 0 \\
0 & 0 & 0 & -1 \\
1 & 0 & 0 & 0 \\
0 & 1 & 0 & 0
\end{array}
\right]
\label{omginv1}
\end{equation} 

Using the BFFT method discussed in the section III, we will find the involutive Hamiltonian for this system as \cite{VPRT}
\begin{eqnarray}
&&\tilde H = \frac {1}{2}\cdot (\bar P_a - \partial_a\bar f(x)\vartheta^{(4)}) (\bar P^a - \partial^a\bar f(x)\vartheta^{(4)}) + \bar V(x) - (\lambda + \vartheta^{(1)})(f(x) + \vartheta^{(2)})
\label{modhm1}
\end{eqnarray}
As mentioned above, all the barred quantities are functions of $x$ and $\vartheta^{(2)}$.
It can be easily verified that the Hamiltonian $\tilde H$ is involutive by computing it's Poisson bracket with modified constraints of the theory.
\begin{eqnarray}
\{\tilde H, \tilde \Phi_i\} = 0
\label{evhmcn}
\end{eqnarray}
where $i = 1,2,3,4$.

\section{Hamiltonian BRST Formalism}
\subsection{Charge and Symmetry }
To construct BRST symmetry for this system, we further extend the theory using Hamiltonian BRST formalism also called BFV formalism \cite{EFGV,IBEF,IABGV}. Here we will present a simplified form of this formalism. 

In the BFV formulation associated with a general class of system with first class constraints, we introduce two canonical set of ghost and anti-ghost fields  $(C^k,{\bar P}_k)$ with ghost number 1 and -1 respectively and $(P^k,{\bar C}_k)$ with ghost number -1 and 1 respectively with Lagrange multiplier fields $(N^k,B_k)$.
These ghost-antighost, corresponding momenta and stuckelberg fields satisfy following super algebra,
\begin{eqnarray}
\{C^k,{\bar P}_l\} = \{P^k,{\bar C}_l\} = \{N^k,B_l\} =\delta^k_l
\label{ghagha}
\end{eqnarray}
Now, the general expression for nilpotent BRST charge, gauge-fixing fermion and BRST invariant Hamiltonian is given by,
\begin{eqnarray}
Q_b = \int dx^N (C^k{\tilde\Phi}_k + P^k B_k)
\label{gnbrch}
\end{eqnarray} 
\begin{eqnarray}
\Psi = \int dx^N({\bar P}_k N^k + \bar C^K \chi_k )
\label{genpsi} 
\end{eqnarray}
 
\begin{eqnarray}
H_U = H_P + H_{BF} - \{Q_b, \Psi\} 
\end{eqnarray}

In BFV formulation the generating functional is independent of gauge fixing fermion \cite{EFGV,IBEF,IABGV,LDFAD,PSENJ}, hence we have liberty to choose it in the convenient form.It is also worth notice that $\chi_k$ is the Hermitian gauge-fixing function with the
identical Grassmann parity as $\tilde\Phi$, and satisfies
\begin{eqnarray}
\det|\{\chi_k,\tilde\Phi_l\}| \neq 0
\end{eqnarray}

Now, we will apply these general results to the particle on surface $V^{(N-1)}$ embedded in $R^N$.  
\begin{eqnarray}
S_{eff} = \int dt \big[ P_a\dot x^a + \Pi_{\vartheta_k} \dot\vartheta^k + B_k{\dot N}^k + {\dot P}^k {\bar C}_k  + {\dot C}^k {\bar P}_k - H_P - H_{BF} + \left[Q_b, \Psi \right] \big] 
\label{eapstk}
\end{eqnarray}
In case of the system described in section IV, the convenient choice is ${\tilde \Phi}_k$ calculated in eqn(\ref{modcn1}).

Our choice of gauge condition $\chi_k$ in this case is $\Phi_k$ in eqn(\ref{cnoth}).

and $\tilde H = H_P + H_{BF}$ is taken as the BFFT modified Hamiltonian in eqn(\ref{modhm1}).

The canonical brackets for all dynamical variables are written as
\begin{eqnarray}
&&[x^a, P_b] = \delta^a_b; \quad [\vartheta_i, \Pi_\vartheta^j ] = \delta^j_i; \nonumber\\ &&\{\bar C_a, \dot C^b\} = i\delta^b_a; \quad\{C^a, \dot{\bar C}_b\} = - i
\delta^a_b
\label{cbfadb}
\end{eqnarray} 
Nilpotent BRST transformation corresponding to this action is constructed using the relation $s_b\Gamma = - [Q_b, \Gamma]_{\pm}$ which is related to infinitesimal BRST transformation as $\delta_b \Gamma = s_b \Gamma \delta \Lambda$. Here $\delta \Lambda$ is infinitesimal BRST parameter. Here $-$ sign is for bosonic and $+$ is for fermionic variable. The BRST transformation for the particle on a Riemann surface is,
\begin{eqnarray} 
s_b N^k &=& P^k, \quad s_b \bar P^k = \tilde \Phi^k\nonumber\\
s_b \bar C^k &=& B^k,\quad s_b C^k = s_b B^k = s_b P^k = 0
\label{brtrf}
\end{eqnarray}
One can easily verify that these transformations are nilpotent.

Using the expressions for $Q_b$ and $\Psi$, Effective action (\ref{eapstk}) is written as
\begin{eqnarray}
S_{eff} = \int dt \big[ P_a{\dot x}^a + \Pi_{\vartheta_k} \dot\vartheta^k + B_k{\dot N}^k + {\dot P}^k {\bar C}_k  + {\dot C}^k {\bar P}_k - \tilde H - P^k{\bar P}_k + N_k \tilde\Phi^k + B_k\chi^k + {\bar C_k}C^k \big ]
\label{eleps}
\end{eqnarray}
and the generating functional for this effective theory is represented as 
\begin{eqnarray}
Z_\psi &=& \int [D \phi]  \exp \left[iS_{eff} \right]
\label{gnf}  
\end{eqnarray}
The Liouville measure $ D\phi =\prod_i d\xi_i $, where $\xi_i$ are all dynamical variables $(P_a, x^a, \Pi_{\vartheta_k}, \vartheta^k, B_k, N^k, {\bar C^k}, P_k, C_k, {\bar P^k} )$ of the theory.
Now integrating this generating functional over $P$ and $\bar P$, we get 
\begin{eqnarray}
&&{Z_\psi} =  \int D \phi' \exp \big[i\int dt \big[P_a {\dot x}^a  + \Pi_{\vartheta^k} \dot\vartheta^k + B_k\dot N^k + \dot C_k{\dot {\bar C}^k} - \tilde H + N_k\tilde \Phi^k - C^k \bar C_k - B_k\chi^k \big] \big]
\label{gfaioppb}    
\end{eqnarray}
where $D\phi'$ is the path integral measure for effective theory when integration over fields $P$ and $\bar P$ are carried out.
Further integrating over $B^k$ we obtain an effective generating functional as
\begin{eqnarray}
&&{Z_\psi} = \int D \phi'' \exp \big[i\int dt \big[ P_a {\dot x}^a  + \Pi_{\vartheta_k} \dot\vartheta^k + \dot C_k{\dot {\bar C}^k} - \tilde H + N_k\tilde\Phi^k - C^k \bar C_k - \frac{\{\dot{N_k} - \chi_k\}^2}{2}  \big] \big]
\label{gfaiopl}    
\end{eqnarray}
where $D\Phi''$ is the path integral measure corresponding to all the dynamical variables involved in the effective action. The BRST symmetry transformation for this effective theory is written as 
\begin{eqnarray} 
s_b N^k &=& \dot{C^k},\quad s_b \bar C^k = - \dot{N^k} - \chi^k \nonumber\\
s_b P^k &=& s_b C^k = s_b B^k = 0 
\label{btrf}
\end{eqnarray}

We know from the literature that BRST charge is nilpotent in nature. Also its operation on the states of Hilbert space will give us the physical subspace of the system. 
\begin{eqnarray}
{Q_{\mathrm{BRST}}} |\mathrm{phys}\rangle = 0, \quad |\mathrm{phys}\rangle \neq {Q_{\mathrm{BRST}}} |....\rangle
\label{bcos}
\end{eqnarray}
which can be written as
\begin{eqnarray}
iC_k\tilde\Phi^k|\mathrm{phys}\rangle  = 0, \quad i\dot{\bar C}_k N^k |\mathrm{phys}\rangle  = 0
\label{bcost}
\end{eqnarray}
This implies that the first class constraints of the system anihilates the physical subspace of the total Hilbert sapce of the system.

\subsection{Canonical BRST Quantization}
The BRST extended action is given by eqn(\ref{eapstk}).As we know, variation of $S$ will give boundary conditions. 
To covariantly quantize this system, we will now Fourier decompose the BRST charge \cite{MKKOSH,SH}.
\begin{eqnarray}
{\tilde\Phi}^k(x,\vartheta,t) &=& \frac{1}{2\pi}\sum_{n = 0}^{\infty}({\tilde\Phi}^k_n e^{-int} + {\tilde\Phi}_n^{k^\dagger} e^{int})\nonumber\\
{P^k(x,t)} &=& \frac{1}{2\pi}\sum_{n = 0}^{\infty}(P^k_n e^{-int} + {P^k_n}^\dagger e^{int})\nonumber\\
C^k(x,t) &=& \sum_{n = 0}^{\infty}({C(x)}^k_n e^{-int} + {{C(x)}^k_n}^\dagger e^{int})\nonumber\\
{\mathcal P^k(x,t)} &=& \sum_{n = 0}^{\infty}({\mathcal P(x)}^a_n e^{-int} + {{\mathcal P(x)}^a_n}^\dagger e^{int})\nonumber\\
B^k(x,t) &=& \frac{1}{2\pi}\sum_{n = 0}^{\infty}({B}^k_n e^{-int} + {{B}^k_n}^\dagger e^{int})
\label{mdepf}  
\end{eqnarray}
Here the commutation relations between these variables is defined as in eqn(\ref{cbfadb}).
Putting these mode expansions in eqn(\ref{gnbrch}) and simplifying, we can easily achieve the operator form of BRST charge.

Applying this charge on the states of total Hilbert space will give us physical subspace conditions.

\section{Examples of $(N - 1)$ Dimensional Embedding in $R^N$} \label{Examples of (N - 1) Dimensional Embedding in R^N}

As an example of ${N-1}$ Dimensional Embedding in $R^N$ we will discuss particle on torus model.

\subsection{Particle on Torus}
 Particle on torus \cite{STH,DXQLXZ,VPBM,RK,ASSG} is a two dimensional surface embedded in three dimensional space. It has been studied as a toy model for different type of field theories. Here, we will discuss all the important results developed for general system in section IV for this case. 
Lagrangian for a particle constrained to move on the surface of torus of radius $r$ is
\begin{eqnarray}
L = \frac{1}{2} m {\dot r}^2+\frac{1}{2} m r^2 {\dot \theta}^2+\frac{1}{2}m (b+r \sin\theta)^2{\dot\phi}^2 + \lambda (r-a)
\label{lagft}
\end{eqnarray}
where $(r,\theta,\phi)$ are toroidal co-ordinates. Their relation  with Cartesian coordinates can be written as,
\begin{equation}
x = (b+r\sin\theta)\cos\phi,
\quad y = (b+r\sin\theta)\sin\phi,
\quad z = r\cos\theta
\end{equation}
and $\lambda$ is the Lagrange multiplier.
Here we have considered a torus with axial circle in the $x - y$ plane centered at the origin, of radius b, having a circular cross section of radius r. The angle $\theta$ ranges from 0 to $2\pi$, and the angle $\phi$ from 0 to $2\pi$.
                       
The canonical Hamiltonian corresponding to the Lagrangian in eqn(\ref{lagft})is then written as,  
\begin{eqnarray}
H = \frac{p^2_r}{2m}+\frac{p^2_\theta}{2mr^2}+\frac{p^2_\phi}{2m(b+r\sin\theta)^2} -\lambda (r-a)
\end{eqnarray}
where $p_r$, $p_\theta, p_\phi$ and $p_\lambda$ are the canonical momenta conjugate to the coordinate $r$, $\theta$ $\phi$ and  $\lambda$ respectively, given by
\begin{eqnarray}
&&p_r = m\dot r, \quad p_{\theta} = m r^2\dot\theta \nonumber\\ 
&& p_\phi = m (b+r\sin\theta)^2\dot\phi, \quad p_\lambda \approx 0
\end{eqnarray}
Here $p_\lambda$ is the primary constraint of the theory. 
fter inclusion of primary constraint our new Hamiltonian has the form
\begin{eqnarray}
H_T = \frac{p^2_r}{2m}+\frac{p^2_\theta}{2mr^2} + \frac{p^2_\phi}{2m(b+r\sin\theta)^2} - \lambda (r-a) + u p_\lambda
\end{eqnarray}
where $u$ is a Lagrange multiplier.
Now, using Dirac's method of Hamiltonian analysis, we will calculate all the possible constraints of the theory. 
\begin{eqnarray}
&&{\dot p}_\lambda = \{ p_\lambda, H_T \}_P = (r-a) \approx 0\nonumber\\
&&{\ddot {p}}_\lambda = \{ (r-a), H_T \}_P = \frac {p_r}{m}  \approx 0 \nonumber\\
&&{p_\lambda}^{(3)} = \{\frac {p_r}{m}, H_T \}_P = \frac{1}{m}\{\frac{p^2_\theta}{mr^3} + \frac{p^2_\phi \sin\theta}
{m(b + r\sin\theta)^2} \nonumber\\ &&+ \lambda \}\approx 0 
\end{eqnarray}
$(P_\lambda)^{(4)}$ will vanish and the value of $u$ will be determined from it.
All the constraints can be written as,
\begin{eqnarray}
&&\Phi_1 = p_\lambda \approx 0 \nonumber\\
&&\Phi_2 = f(x) = (r-a)\approx 0 \nonumber\\
&&\Phi_3 = Df(x) = \frac {p_r}{m} \approx 0 \nonumber\\
&&\Phi_4 = D^2 f(x) - {\partial_b}(V - \lambda f(x))\cdot \partial^b f(x) \nonumber\\ &&= \frac{1}{m}\{\frac{p^2_\theta}{mr^3} + \frac{p^2_\phi \sin\theta}{m(b + r\sin\theta)^2} + \lambda \} \approx 0
\label{cnopt}
\end{eqnarray}
Now, the Poisson brackets between the constraints have following values,
\begin{eqnarray}
&&\{\Phi_1, \Phi_4 \}_P = - \partial_a f(x)\cdot \partial^a f(x) = - \frac{1}{m} \nonumber\\
&&\{\Phi_2, \Phi_3 \}_P = \partial_a f(x)\cdot \partial^a f(x) = \frac{1}{m} \nonumber\\
&&\{\Phi_2, \Phi_4 \}_P = P^a\partial_a (\partial_b f(x)\cdot \partial^b f(x)) = P^a(\partial_a \alpha) = 0 \nonumber\\
&&\{\Phi_3, \Phi_4 \}_P = 2\partial_a (D f(x))\cdot \partial^a (D f(x)) - \partial_a f(x)\nonumber\\ &&\cdot \partial^a (D^2 f(x) - {\partial_b}\Phi\cdot \partial^b f(x)) \nonumber\\ &&= \frac{3}{m^3}\{\frac{p^2_\theta}{r^4} + \frac{p^2_\phi \sin^2\theta}
{(b + r\sin\theta)^4} \} = - \gamma 
\end{eqnarray}
Thus the matrix $\Delta_{ab}$ between the constraints has the form
\begin{equation}
\Delta_{ab} = \{\Phi_a, \Phi_b\}_P =
\small
\left[
\begin{array}{cccc}
0 & 0 & 0 & - \frac{1}{m} \\
0 & 0 & \frac{1}{m} & 0 \\
0 & -\frac{1}{m} & 0 & \frac{3}{m^3}\{\frac{p^2_\theta}{r^4} + \frac{p^2_\phi \sin^2\theta}
{(b + r\sin\theta)^4} \}  \\
\frac{1}{m}& 0 & - \frac{3}{m^3}\{\frac{p^2_\theta}{r^4} + \frac{p^2_\phi \sin^2\theta}
{(b + r\sin\theta)^4} \}  & 0
\end{array}
\right]
\end{equation}
As all the constraints of the theory (\ref{cnopt}) are second class, we will follow the method of section IV and introduce four possible fields $\vartheta^1, \vartheta^2, \vartheta^3, \vartheta^4$ corresponding to each constraint. Relation between these fields will provide us possible value of $\omega^{ab}$.
Our choice of Poisson bracket between the fields will be same as  one taken for the general case.
Hence the matrix $\omega^{ab}$ will have the form of eqn(\ref{omgab1}).

Using the matrix $\omega^{ab}$ and the matrix $\Delta_{ab}$ between the constraints in the eqn(\ref{exton}), one can find many possible value of matrix $X_{ab}$. 

Now, applying the results developed in section IV we can calculate the modified constraints as
\begin{eqnarray}
\tilde\Phi_1 &=& p_\lambda - \vartheta^{(3)}\nonumber\\
\tilde\Phi_2 &=& (r - a) + \vartheta^{(2)} \nonumber\\
\tilde\Phi_3 &=& \frac{1}{m}(p_r - \vartheta^{(4)}) \nonumber\\
\tilde\Phi_4 &=& \frac{1}{m}\{\frac{p^2_\theta}{m(r + \vartheta^{(2)})^3} + \frac{p^2_\phi \sin\theta}{m(b + (r+\vartheta^{(2)})\sin\theta)^3} + \lambda - \theta^{(1)}\}
\end{eqnarray}
The Poisson bracket between these modified constraints vanishes which shows that modified constraints are involutive. Hence we have converted the second class constraints of the theory into first class.

Now, we will construct first class Hamiltonian for this system using the results in section V.

The total involutive Hamiltonian for this system will take the form as \cite{VPRT},
\begin{eqnarray}
&&\tilde H = \frac{(p_r - \vartheta^{(4)})^2}{2m}+\frac{p^2_\theta}{2m(r + \vartheta^{(2)})^2} + \frac{p^2_\phi}{2m(b+(r+ \vartheta^{(2)})\sin\theta)^2} - (\lambda + \vartheta^{(1)})(r + \vartheta^{(2)}-a) 
\end{eqnarray}
It can be easily verified that the Hamiltonian $\tilde H$ is involutive by computing it's Poisson bracket with modified constraints of the theory.
\begin{eqnarray}
\{\tilde H, \tilde \Phi_i\} = 0
\end{eqnarray}
where $i = 1,2,3,4$.

BRST charge for this first class system can be written using above expression, as
\begin{eqnarray}
Q_b = iC^i\tilde\Phi_i + iP^iB_i
\label{brchtr}
\end{eqnarray}

and corresponding BRST symmetry transformation can be written as
\begin{eqnarray}
s_b N^i &=& P^i, \quad s_b \bar P^i = \tilde \Phi^i\nonumber\\
s_b \bar C &=& B^i,\quad s_b C = s_b P^a = s_b B^i = s_b P^i = 0
\label{brsytr} 
\end{eqnarray}

This shows that result obtained above is true for any $(N-1)$ dimensional surface embedded in $R^N$.

\section{Batalin - Vilkovisky Quantization}
We will perform the quantization of system described above along the field-antifield formalism \cite{IABGAV,IBGV,JGJPSS} for BFFT system discussed in ref \cite{RART}. To do so we will introduce antifields 
$\phi^{\star}_A = ( x^{\star}_\mu, \vartheta^{\star}_\nu, N^{\star}_\nu, C^{\star}_\nu )$ corresponding to the fields $\phi^A = (x^\mu, \vartheta^\nu, N^\nu, C^\nu)$. Here, fields $x^\mu, \vartheta^\nu$ and $N^\nu$ are bosonic and have ghost number zero. The ghosts $C^\nu$ are fermionic and have ghost number one. The corresponding anti-fields have opposite grassmanian parity and ghost number given by minus the ghost number of the corresponding field minus one.
Action in terms of fields and antifields is written as
\begin{eqnarray}
&&S = S_0 + \int dt \big[ x^{\star}_\mu \{ x^\mu, {\tilde\Phi}_\nu \} c^\nu + {\vartheta}^{\star}_\varphi \{ \vartheta^\varphi, {\tilde\Phi}_\nu \} c^\nu + N^{\star}_\nu {\dot C}^\nu \big ]
\label{acbff}
\end{eqnarray}
where $S_0$ is defined as
\begin{eqnarray}
S_0 & = &\int dt \big[ P_\mu{\dot x}^\mu + \Pi_\nu {\dot\vartheta}^\nu - N^\nu {\tilde\Phi}_\nu - {\tilde H}\big ]
\label{ginac}
\end{eqnarray}
Here ${\tilde\Phi}$ are the modified constraints in eqn(\ref{modcn1}) and ${\tilde H}$ is the modified Hamiltonian in eqn(\ref{modhm1}).
Now, this action satisfies the classical master equation 
\begin{eqnarray}
\frac{1}{2}(S,S) = 0
\label{clmeq}
\end{eqnarray}
where the antibracket between any two quantities $X[\phi, \phi^{\star}]$ and $Y[\phi, \phi^{\star}]$ is defined as 
\begin{eqnarray}
(X,Y) = \frac{\delta_r X}{\delta \phi^A}\frac{\delta_l Y}{\delta \phi^{\star}_A} - \frac{\delta_r X}{\delta\phi^{\star}_A}\frac{\delta_l Y}{\delta \phi^A }
\label{antib}
\end{eqnarray}
Here we assume the de Witt's notation of sum and integration over intermediary variables, when necessary.
In the BV formalism, the BRST differential is introduced using the relation
\begin{eqnarray}
sX = (X,S)
\label{brdif}
\end{eqnarray} 
for any local functional $X[\phi, \phi^{\star}]$. Due to classical master equation and Jacobi identity, s is nilpotent. So, the BV action satisfying the master equation is equivalent to BRST invariance.

To fix a gauge, we need to introduce trivial pairs ${\bar C}_\nu, P_\nu$ as new fields and the corresponding antifields $\bar {C^{\star}_\nu}, P^{\star}_\nu$, as well as a gauge-fixing fermion $\Psi$. The antifields are eliminated by choosing $\phi^{\star}_A = \frac{\partial \Psi}{\partial \phi^A}$. We can choose the form of $\Psi$ as
\begin{eqnarray}
\Psi = {\bar C}_\nu \vartheta^\nu
\label{gffer}
\end{eqnarray} 
Other possible choices can also be made. It is also necessary to extend the field-antifield action to a nonminimal action,
\begin{eqnarray}
S \rightarrow S_{nm} = S + \int dt P_\nu {\bar C}^{{\star}\nu}
\label{nnmac} 
\end{eqnarray}
in order to implement the gauge fixing introduced by $\Psi$. The gauge-fixed generating functional is then defined as
\begin{eqnarray}
Z_\Psi = \int [d\phi^A ][d\omega ]^{-\frac{1}{2}} [df]^{-\frac{1}{2}} \exp {\frac{i}{\hbar}S_{nm}\big[\phi^A, \phi^{\star}_A = \frac{\partial \Psi}{\partial \phi^A} \big ]}
\label{vacfn}    
\end{eqnarray}
In general, if the classical field-antifield action $S$ is replaced by some quantum action $W$ expressed as a local functional of fields and antifields and satisfy the so-called quantum master equation
\begin{eqnarray}
\frac{1}{2}(W,W) - i{\hbar}\Delta W = 0
\label{qnmeq}
\end{eqnarray}
then the gauge symmetries are not obstructed at quantum level. 
Here $\Delta$ is a potentially singular operator which is defined as
\begin{eqnarray}
\Delta \equiv (\frac{\delta_r}{\delta \phi^A})(\frac{\delta_l}{\delta \phi^{\star}_A})
\label{}
\end{eqnarray}
and it was assumed that $W$ can be expanded in powers of $\hbar$ as
\begin{eqnarray}
W[\phi^A,\phi^{\star}_A] = S[\phi^A,\phi^{\star}_A] + \sum_{p = 1}^{\infty}{\hbar}^p M_p[\phi^A,\phi^{\star}_A]
\label{qnact}
\end{eqnarray}
The first two term of the quantum master equation (\ref{qnmeq}) are
\begin{eqnarray}
(S,S) = 0\nonumber\\
(M_1,S) = i\Delta S
\label{qnaeq}
\end{eqnarray}
If $\Delta S$ is non-zero and gives a nontrivial result, then there exists some $M_1$ expressed in terms of local fields such that (\ref{qnaeq}) is satisfied.
Using cohomological arguments, it can be shown that the quantum master equation, for first order systems with pure second class constraints converted with the use of the BFFT procedure, can always be solved. BRST transformations for the BFFT converted system can be written as
\begin{eqnarray} 
&&s_b N^\nu = {\dot C}^\nu, \quad s_b C^\nu = 0, \quad s_b {\bar C}^\nu = P^\nu, \quad s_b P^\nu  = 0\nonumber\\
&&s_b x^{\star}_\mu = -\frac{\partial S}{\partial x^\mu}, \quad s_b \vartheta^{\star}_\nu = -\frac{\partial S}{\partial {\vartheta}^\nu},\quad s_b N^{\star}_\nu = {\tilde\Phi}_\nu, \nonumber\\&& s_b C^{\star}_\nu = - x^{\star}_\mu \{ x^\mu, {\tilde\Phi}_\nu \} - {\vartheta}^{\star}_\varphi \{ \vartheta^\varphi, {\tilde\Phi}_\nu \} - {\dot N}^\star, 
\quad s_b \bar{C^{\star}_\nu} = 0,\quad s_b {\bar P}^{{\star}\nu} =  {\bar C}^{{\star}\nu} 
\label{bvbrst}
\end{eqnarray}
These symmetry transformations are same as the one obtained in (\ref{btrf}). It can be shown, on the basis of argument given in ref {\cite{RART}} that enlarged symmetries due to compensating fields (BFFT variables) are not anomalous. These fields plays non-trivial role at the quantum level because the existence of a counter-term modify expectation values of relevant physical quantities.

\section{Conclusion}
We have for the first time investigated BRST symmetry for a particle moving in a curved space $V_{(N - 1)}$ embedded in a Euclidean space $R_N$ in both Hamiltonian and Lagrangian formalism. Using the Dirac's constraints analysis, we have calculated all the constraints of the system. Using the BFFT technique, second class constraints are converted into first class constraints and corresponding first class Hamiltonian is constructed. In the limit of $\vartheta \rightarrow 0$ the constraints and Hamiltonians return to original second class one. Now, using BFV technique we have constructed BRST charge and corresponding BRST invariant action. We have shown that, the action of BRST charge on the state of total Hilbert space will give physical subspace of the system. We have constructed BRST operator using mode expansion technique. This is the first time in the literature that BFFT Abelianization and BRST symmetry is constructed for a nonlinear second class systems explicitly. We have also discussed simple example (particle on torus) of this kind of system. At the end we have discussed Batalin - Vilkovisky quantization of this system based on BFFT formalism. As this model is a toy model for a wide class of physical systems, the results obtained here will be highly useful in studying these systems. The more general case of BRST quantization of embedding $V_L$ in Euclidean space $R_N$ where $1\leq L < N$  will be discussed in the next part of the paper \cite{VKP}.


\section{References}\label{References}
\begin{thebibliography}{}
\bibitem{LLSKOT} H. E. Lin, W. C. Lin and R. Sugano, Nucl. Phys. B 16, 431 (1970).
\bibitem{RS} R. Sugano, Prog. Theor. Phys. 46, 297 (1971).
\bibitem{TKM} T. Kimura, Prog. Theor. Phys. 46, 1261 (1971).
\bibitem{TKRS} T. Kimura and R. Sugano, Prog. Theor. Phys. 47, 1004 (1972).
\bibitem{TORS1} T. Ohtani and R. Sugano, Prog. Theor. Phys. 47, 1704 (1972).
\bibitem{TORS2} T. Ohtani and R. Sugano, Prog. Theor. Phys. 50, 1715 (1973).
\bibitem{TKTORS} T. Kimura, T. Ohtani and R. Sugano, Prog. Theor. Phys. 48, 1395 (1972).
\bibitem{MOHS} M. Omote and H. Sato, Prog. Theor. Phys. 47, 1367 (1972).
\bibitem{FSTK} K. Fujii, K-I. Sato, N. Toyota and A. P. Kobushukin, Phys. Rev. Lett. 58, 651 (1987).
\bibitem{KFAK} K. Fujii, A. Kobushukin, K-I. Sato and N. Toyota, Phys. Rev. D 37, 3663 (1988). 
\bibitem{TK} T. Kawai, Foundation of Phys. 143 (1975).
\bibitem{HKTK} H. Kamo and T. Kawai, Nucl. Phys. B 81, 349 (1974). 
\bibitem{JGANMO} J. L. Garvais and A. Neveu, Phys. Rep. 23, 237 (1976).
\bibitem{MO} M. Omote, Nucl. Phys. B 120, 325 (1977).
\bibitem{BS} B. Sakita and K. Kikkawa, Quantum Mechanics of Many Degrees of Freedom Basing on Path Integral, Iwanami, Tokyo (1986).
\bibitem{BDMOHS} B. S. Dewitt, Phys. Rev. 85, 653 (1952). 
\bibitem{VASFMR} V. De Alearo, S. Fubini, G. Furlan and M. Roncadelli, Nucl. Phys. B 296, 402 (1988).
\bibitem{NBD} N. D. Birrell and P.C.W. Davies, Quantum fields in curved space, Cambridge University Press, Cambridge (1982).
\bibitem{LPDT} L. Parker and D. Toms., Quantum Field Theory in Curved Spacetime, Cambridge University Press, Cambridge (2009).
\bibitem{IBMGSL1} I. A. Batalin, M. A. Grigoriev and S. L. Lyakhovich, Theoretical and Mathematical Physics, 128, 1109 (2001).
\bibitem{IBMGSL2} I. A. Batalin, M. A. Grigoriev and S. L. Lyakhovich, J. Math. Phys. 46, 072301 (2005).
\bibitem{IGSLAS} I.V. Gorbunov, S.L. Lyakhovich, A.A. Sharapov, J. Geom. and Phys. 53, 98 (2005).
\bibitem{OFK} N. Ogawa, K. Fujii, A. Kobushukin, Prog. Theo. Phys. 83, 894 (1990).
\bibitem{OFCK} N. Ogawa, K. Fujii, N. Chepilko and A. Kobushkin, Prog. Theo. Phys., 85, 1189 (1991).
\bibitem{NOG1} N. Ogawa, Phys. Rev. D 62, 085023 (2000).
\bibitem{MIYN} M. Ikegami, Y. Nagaoka, S. Takagi and T. Tanzawa, Prog. of Theor. Phys. 88, 229 (1992).
\bibitem{STTT} S. Takagi and T. Tanzawa, Prog. of Theor. Phys. 87, 561 (1992). \bibitem{CFFM} C. Filgueiras and F.Moraes, Annals Phys.323, 3150 (2008).
\bibitem{NOG2} N. Ogawa, Phys. Rev. E 81, 061113 (2010).
\bibitem{FOUC} K. Fujii, N. Ogawa, S. Uchiyama and N. Chepilko, Int. J. of Mod. Phys. A 12, 5235 (1997).
\bibitem{NCKFAK} N. Chepilko, K. Fujii, and A. Kobushkin, Phys. Rev. D 44, 3249 (1991).
\bibitem{ASA} A. Saa, Class and Quan. Grav. 14, 385 (1997).
\bibitem{DLLHQL} D. Lian, L. Hu and Q. Liu, Ann. Phys. 530, 1700415 (2018).
\bibitem{NOG3} N. Ogawa, Mod. Phys. Lett. A 12, 1583 (1997).
\bibitem{NOG4} N. Ogawa, arxiv:hep-th/9703181.
\bibitem{NOMN} N. Okamoto and M. Nakamura, Prog. Theo Phys. 96, 235 (1996).
\bibitem{DMOFGK} C. Destri, P. Maraner and E. Onofri, II Nuovo Cimento A (1965-1970), 107, 237 (1994).
\bibitem{AFHGPK} A. Foerster, H. O. Girotti and P. S. Kuhn, Phys. Lett. A 195, 301 (1994).
\bibitem{AVG} A. V. Golovnev, Rept. Math. Phys. 6, 459 (2009).
\bibitem{KFNO} K. Fujii, N. Ogawa, Prog. Theo. Phys. 89, 575 (1993).
\bibitem{MNNOHM1} M. Nakamura, N. Okamoto and H. Minowa, arXiv:hep-th/9710232.
\bibitem{MNNOHM2} M. Nakamura, N. Okamoto and H. Minowa, II Nuovo Cimento B (1971-1996) 111, 521 (1996). 
\bibitem{MN} M. Nakamura, arXiv:1503.06541. 
\bibitem{PM1} P. Maraner, J. Phys. A 28, 2939 (1995). 
\bibitem{PM2} P. Maraner, Annals Phys. 246, 325 (1996). 
\bibitem{NCAR1} N. Chepilko and A. Romanenko, Eur. Phys. J. C 21, 369 (2001).
\bibitem{NCAR2} N. Chepilko and A. Romanenko, Eur. Phys. J. C 21, 587 (2001).
\bibitem{NCAR3} N. Chepilko and A. Romanenko, Eur. Phys. J. C, 21, 757 (2001).
\bibitem{SOCSRT} S. L. de Oliveira, C. M. B. Santos and R. Thibes, Braz. J. Phys. B, 50 (2020) 480.
\bibitem{BRS1} C. Becchi, A. Rouet and R. Stora, Phys. Lett. B 52, 344 (1974).
\bibitem{BRS2} C. Becchi, A. Rouet and R. Stora, Commun. Math. Phys. 42, 127 (1975).
\bibitem{BRS3} C. Becchi, A. Rouet and R. Stora, Ann. Phys. 98, 287 (1976). 
\bibitem{IVT} I. V. Tyutin, Lebedev Report N FIAN, 39 (1975), arXiv:0812.0580 (2008).
\bibitem{PAMD1} P. A. M. Dirac, Can. J. Math. 2, 129 (1950).
\bibitem{JAPB} J. L. Anderson and P. G. Bergmann, Phys. Rev. 83, 1018 (1951).
\bibitem{PAMD2} P. A. M. Dirac, Lectures on Quantum Mechanics, Yeshiva University, New York (1964).
\bibitem{KS} K. Sundermeyer, Constrained Dynamics, Lecture notes in Physics, vol. 169, Springer, Berlin (1982).
\bibitem{MHCT} M. Henneaux and C. Teitelboim, Quantization of Gauge System, Princeton University Press, Princeton (1992).
\bibitem{MH} M. Henneaux, Phys. Rep. 126, 1 (1985).
\bibitem{LFSS} L. D. Faddeev and S. L. Shatashvili, Phys. Lett. B 167, 225 (1986).
\bibitem{IBEF1} I. A. Batalin and E. S. Fradkin, Phys. Lett. B 180, 157 (1986). 
\bibitem{IBEF2} I. A. Batalin and E. S. Fradkin, Nucl. Phys. B 279, 514 (1987).
\bibitem{IBEFTF} I. A. Batalin, E. S. Fradkin, and T. E. Fradkina, Nucl. Phys. B 314, 158 (1989)[Erratum: Nucl. Phys. B 323, 734 (1989)]. 
\bibitem{IBIT} I. A. Batalin and I. V. Tyutin, Int. J. Mod. Phys. A, 6, 3255 (1991).
\bibitem{EERM} E. S. Egorian and R. P. Manvelyan, Theor. Math. Phys. 94, 173 (1993).
\bibitem{RBJBN} R. Banerjee and J. Barcelos-Neto, Nucl. Phys. B 499, 453 (1997).
\bibitem{CBAPMT} C. Burdik, A. Pashnev and M. Tsulaia, Mod. Phys. Lett. A 16, 731 (2001).
\bibitem{IBVKAR} I. L. Buchbinder, V. K. Krykhtin and A. A. Reshetnyak, Nucl. Phys. B 787, 211 (2007).
\bibitem{ILBAAR} I. L. Buchbinder and A. A. Reshetnyak, Nucl. Phys. B 862, 270 (2012).
\bibitem{AAR} A. A. Reshetnyak, JHEP 1809, 104 (2018).
\bibitem{IBAR} I. L. Buchbinder and A. A. Reshetnyak, Phys. Lett. B 820, 136470 (2021).
\bibitem{EFGV} E. S. Fradkin and G. Vilkovisky, Phys. Lett. B 55, 224 (1975).
\bibitem{IABGV} I. A. Batalin and G. Vilkovisky, Phys. Lett. B 69, 309 (1977).
\bibitem{IBEF} I. A. Batalin and E. S. Fradkin, Phys. Lett. B 122, 157 (1983).
\bibitem{LDFAD} L. D. Faddeev, Theor. Math. Phys. 1, 1 (1970).
\bibitem{PSENJ} P. Senjanovic, Ann. Phys. 100, 277 (1976) [Erratum: Ann. Phys. 209 (1991) 248].
\bibitem{MKKOSH} M. Kato and K. Ogawa, Nucl. Phys. B 212, 443 (1983).
\bibitem{SH} S. Hwang, Phys. Rev. D 28, 2614 (1983).
\bibitem{STH} S. T. Hong, Mod. Phys. Lett. A 20, 1577 (2005).
\bibitem{VPBM} V. K. Pandey and B. P. Mandal, Adv. High Energy Phys. 2017, 6124189 (2017). 
\bibitem{DXQLXZ} D. M. Xuna, Q.H. Liu and X.M. Zhu, Annals Phys. 338 , 123 (2013).
\bibitem{RK} R. Kumar, EPL 106, 51001 (2014) [erratum: EPL 108, 59902 (2014)].
\bibitem{ASSG} Anjali S. and S. Gupta, EPL 135, 11002 (2021).
\bibitem{IABGAV} I. A. Batalin and G. A. Vilkovisky, Phys. Lett. B 102, 27 (1981).
\bibitem{IBGV} I. A. Batalin and G. A. Vilkovisky, Phys. Rev. D 28, 2567 (1983)[Erratum-ibid: Phys. Rev. D 30, 508 (1984)].
\bibitem{JGJPSS} J. Gomis, J. Paris and S. Samuel, Phys. Rep., 259, 1 (1995).
\bibitem{RART} R. Amorim and R. Thibes, J. Math. Phys., 40, 5306 (1999).
\bibitem{VKP} V. K. Pandey, Hamiltonian and Lagrangian BRST quantization in Riemann Manifold II (in preparation)
\bibitem{VPRT} V. K. Pandey and R. Thibes, arxiv: 2103.05626[hep-th]
\bibitem{VKP2} V. K. Pandey, arxiv:2007.00452[hep-th]
\end{thebibliography}
\end{document}